\documentstyle[11pt,IAU207_pasp,twoside,psfig]{article}
\def\edcomment#1{\iffalse\marginpar{\raggedright\sl#1\/}\else\relax\fi}
\reversemarginpar

\begin{document}
\title{Tidally-induced Super Star Clusters in M82}
 \author{R. de Grijs}
\affil{Institute of Astronomy, Univ.  of Cambridge, Madingley Road,
Cambridge CB3 0HA, UK}
\author{R.W. O'Connell}
\affil{Astronomy Dept., Univ.  of Virginia, P.O.  Box 3818,
Charlottesville, VA 22903, USA}
\author{J.S. Gallagher, {\sc iii}}
\affil{Astronomy Dept., Univ. of Wisconsin, 475 N. Charter Str., Madison,
WI 53706, USA}

\begin{abstract}
Using new {\sl HST} imaging, we identify a large, evolved system of
super star clusters in a disk region just outside the starburst core in
the prototypical starburst galaxy M82, ``M82 B.'' This region has been
suspected to be a fossil starburst site in which an intense episode of
star formation occurred over 100 Myr ago, which is now confirmed by our
derived age distribution.  It suggests steady, continuing cluster
formation at a modest rate at early times ($> 2$ Gyr ago), followed by a
concentrated formation episode $\sim 600$ Myr ago and more recent
suppression of cluster formation.  The peak episode coincides with
independent dynamical estimates for the last tidal encounter with M81.
\end{abstract}

\section{M82, the prototypical starburst galaxy}

Observations from radio to X-rays show evidence for a tidally-induced
starburst in the center of M82 (e.g., Telesco 1988) In fact, there is
now evidence that M82 has undergone multiple episodes of intense star
formation (cf.  Marcum \& O'Connell 1996, de Grijs, O'Connell \&
Gallagher 2001).  ``M82 B,'' the fossil starburst region, has exactly
the properties of an {\em evolved} starburst with a similar amplitude to
the active burst (Marcum \& O'Connell 1996).  Thus, M82 is a unique
starburst galaxy, since no other galaxy offers the opportunity to study
two discrete starbursts at such close range! By analogy with the {\sl
HST} results from the active starburst region (O'Connell et al.  1994,
1995), we expected M82 B to have contained a complement of luminous
super star clusters.  It is possible that most of the star formation in
starbursts takes place in the form of such concentrated clusters; in
M82, we do not observe any outside the starburst regions. 

The combination of observations of both the active and the fossil
starburst sites in M82 provides a unique physical environment for the
study of the stellar and dynamical evolution of star cluster systems

\section{The star formation history in M82}

\begin{figure}
\centerline{\vbox{
\psfig{figure=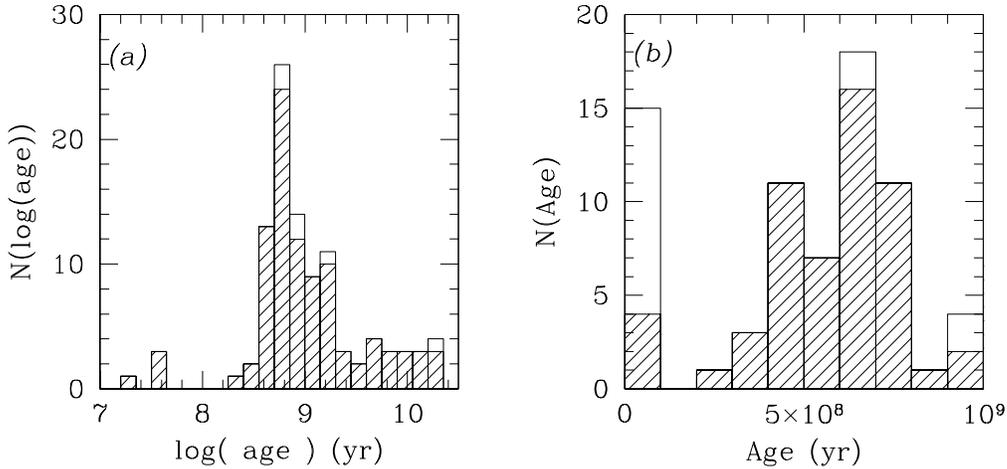,width=14cm}
}}
\vspace*{-7.5cm}
\caption[]{Age distribution of the star clusters in M82 B.  Shaded
histogram: well-determined ages; open histogram: upper or lower limits
only (cf.  de Grijs et al.  2001).}
\end{figure}

Based on {\sl HST/WFPC2} observations of two adjacent fields in M82 B
(in F439W, F555W and F814W), we detected $\sim 100$ (slightly) evolved
star clusters.  The clusters brighter than $V = 22.5$ exhibit a wide
range of ages, from $\sim 30$ Myr to over 10 Gyr.  There is a strong
peak of cluster formation at $\sim 600$ Myr, in either representation in
Fig.  1; very few clusters are younger than 300 Myr. 

Our results suggest steady, continuing cluster formation at a very
modest rate at early times ($> 2$ Gyr ago), followed by a concentrated
formation episode lasting from 400--1000 Myr ago and a subsequent
suppression of cluster formation.  Thus, it appears that the last tidal
encounter between M82 and M81 $\sim$ 500--800 Myr ago had a major impact
on what was probably an otherwise normal, quiescent, disk galaxy.  It
caused a concentrated burst of star formation activity, which decreased
rapidly within a few hundred Myr. 

M82 B has evidently not been affected by the more recent ($< 30$
Myr) starburst episode now continuing in the central regions.

\end{document}